\documentclass[a4paper,fleqn,usenatbib]{mnras}
\usepackage[T1]{fontenc}
\usepackage{ae,aecompl}
\usepackage{graphicx}	
\usepackage{amsmath}	
\usepackage{amssymb}	
\usepackage{amsbsy}
\usepackage{float}
\usepackage{color}
\usepackage{subcaption}

\usepackage{mathrsfs,bm}
\newcommand{\bcdot}{\ensuremath{%
  \mathchoice%
   {\mskip\thinmuskip\lower0.2ex\hbox{\scalebox{1.5}{$\cdot$}}\mskip\thinmuskip}}%
   {\mskip\thinmuskip\lower0.2ex\hbox{\scalebox{1.5}{$\cdot$}}\mskip\thinmuskip}%
   {\lower0.3ex\hbox{\scalebox{1.2}{$\cdot$}}}%
   {\lower0.3ex\hbox{\scalebox{1.2}{$\cdot$}}}%
}

\title[Magnetising the circumgalactic medium]{Magnetising the circumgalactic medium of disk galaxies}

\author[R.~Pakmor et al.]  {R\"udiger~Pakmor$^{1}$\thanks{E-mail: rpakmor@mpa-garching.mpg.de}, Freeke van de Voort$^{1}$, Rebekka Bieri$^{1}$, Facundo A. G\'omez$^{2,3}$,  \newauthor 
       Robert J.~J. Grand$^{1}$, Thomas Guillet$^{4}$, Federico Marinacci$^{5}$, Christoph Pfrommer$^{6}$,  \newauthor Christine M. Simpson$^{7,8}$, Volker Springel$^{1}$
  \vspace*{0.2cm}  \\
  $^1$Max-Planck-Institut f\"{u}r Astrophysik, Karl-Schwarzschild-Str. 1, D-85748, Garching, Germany\\
  $^2$Instituto de Investigaci\'on Multidisciplinar en Ciencia y Tecnolog\'ia, Universidad de La Serena, Ra\'ul Bitr\'an 1305, La Serena, Chile\\
  $^3$Departamento de Astronom\'ia, Universidad de La Serena, Av. Juan Cisternas 1200 Norte, La Serena, Chile\\
  $^4$Department of Physics and Astronomy, Stocker Road, University of Exeter, EX4 4QL, United Kingdom\\
  $^5$Department of Physics and Astronomy, University of Bologna, via Gobetti 93/2, 40129 Bologna, Italy\\
  $^6$Leibniz-Institut f\"{u}r Astrophysik Potsdam (AIP), An der Sternwarte 16, D-14482 Potsdam, Germany\\
  $^7$Department of Astronomy \& Astrophysics, The University of Chicago, Chicago, IL 60637, USA\\
  $^8$Enrico Fermi Institute, The University of Chicago, Chicago, IL  60637, USA\\
}
  
\date{Accepted 2020 August 17. Received 2020 August 17; in original form 2019 November 25}

\pubyear{2019}

\begin{document}

\label{firstpage}
\pagerange{\pageref{firstpage}--\pageref{lastpage}}

\maketitle

\begin{abstract}
The circumgalactic medium (CGM) is one of the frontiers of galaxy formation and intimately connected to the galaxy via accretion of gas on to the galaxy and gaseous outflows from the galaxy. Here we analyse the magnetic field in the CGM of the Milky Way-like galaxies simulated as part of the \textsc{Auriga} project that constitutes a set of high resolution cosmological magnetohydrodynamical zoom simulations. We show that before $z=1$ the CGM becomes magnetised via galactic outflows that transport magnetised gas from the disk into the halo. At this time the magnetisation of the CGM closely follows its metal enrichment. We then show that at low redshift an in-situ turbulent dynamo that operates on a timescale of Gigayears further amplifies the magnetic field in the CGM and saturates before $z=0$. The magnetic field strength reaches a typical value of $0.1\,\mu G$ at the virial radius at $z=0$ and becomes mostly uniform within the virial radius. Its Faraday rotation signal is in excellent agreement with recent observations. For most of its evolution the magnetic field in the CGM is an unordered small scale field. Only strong coherent outflows at low redshift are able to order the magnetic field in parts of the CGM that are directly displaced by these outflows.
\end{abstract}


\begin{figure*}
  \centering
    \begin{subfigure}[t]{.49\textwidth}
    \centering
    \includegraphics[width=\linewidth]{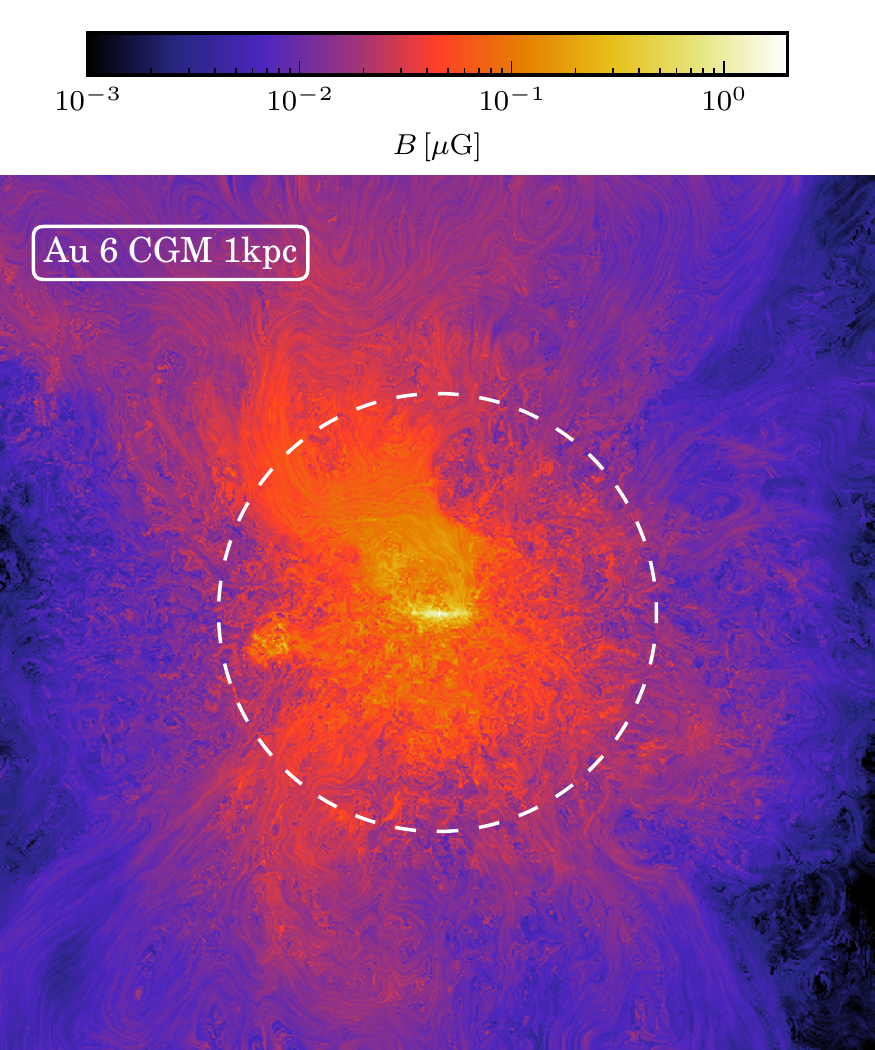}
    \end{subfigure}
    \begin{subfigure}[t]{.49\textwidth}
    \centering
    \includegraphics[width=\linewidth]{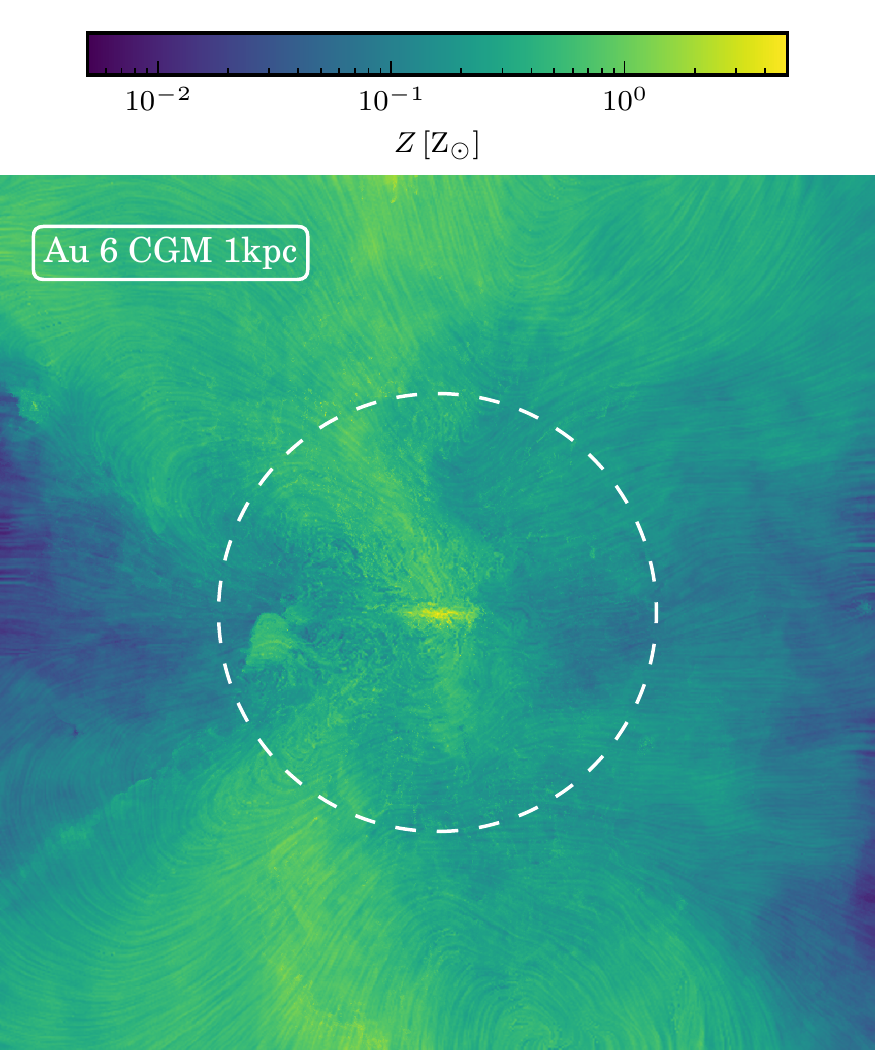}
    \end{subfigure}
  \caption{Edge-on projections of the volume averaged root mean square magnetic field strength (left panel) and mass weighted average metallicity (right panel) at $z=0$ for Au-6-CGM with $1\,\mathrm{kpc}$ minimum spatial resolution in the CGM out to a radius of $400\,\mathrm{kpc}$ \citep{AurigaCGM}. The virial radius at this time is $R_\mathrm{vir}=210\,\mathrm{kpc}$ and is shown by the dashed circle. The width and depth of the of the projections is $4\,R_\mathrm{vir}$. Superimposed on the panels using the line integral convolution method \citep{LIC} are the orientation of the magnetic field in the left panel and the orientation of the velocity field in the right panel.}
  \label{fig:cgm}
\end{figure*}

\section{Introduction}

The CGM plays an important role in the formation and evolution of galaxies and has recently come into the focus of observations as well as numerical simulations. It is intimately connected to the central galaxy, providing the gas reservoir for accretion onto the galaxy, and acting as the repository of gas that is expelled from the galaxy via various feedback channels. Owing to the low gas density of the CGM, it is much harder to observe than the interstellar medium \citep{Tumlinson2017}. New observations \citep[for a review see e.g.][]{Werk2014,Tumlinson2017} and simulations \citep[e.g.][]{AurigaCGM, Peeples2019, Hummels2019, Suresh2019} have made significant progress towards accessing and understanding the CGM, however many fundamental questions remain unanswered.

Modelling the CGM in simulations of galaxies is demanding and has only recently become feasible for Milky Way-like disk galaxies, once high resolution cosmological simulations managed to produce disk galaxies with realistic stellar and gas masses and sizes, which is the minimum requirement to meaningfully model the CGM. It either requires very expensive simulations of cosmological boxes \citep[e.g.][]{TNG50Nelson} or high resolution cosmological zoom simulations \citep[e.g.][]{Guedes2011,Marinacci2014,Auriga,FIRE2}. On top of zoom simulations that simulate a region around a central halo at enhanced constant mass resolution, recent simulations employ additional volume refinement in the halo to improve the spatial resolution specifically in the CGM \citep{AurigaCGM, Peeples2019, Hummels2019}. Independently, idealised small scale simulations have started to probe the conditions relevant for the CGM \citep[see, e.g.][]{Armillotta2016,Armillotta2017,McCourt2018,Sparre2019}. A similar approach has been used to model the magnetic field in the intracluster medium by boosting the spatial resolution to study the amplification of the field by a turbulent dynamo \citep{Vazza2014,Vazza2018}. However, these simulations lacked the resolution to resolve the internal structure of galaxies with sizes of their smallest cells of $13\,\mathrm{ckpc}$ \citep{Vazza2014} and $4\,\mathrm{ckpc}$ \citep{Vazza2018}, which is more than an order of magnitude larger than the softening of recent zoom-in simulations of galaxies, e.g. $250\,\mathrm{cpc}$ for standard resolution simulations of the \textsc{Auriga} project \citep{Auriga}.

Magnetic fields in the CGM are still an unexplored topic. There is so far no good understanding of either the properties of the magnetic field in the CGM and how it evolves with time or the relevance of the magnetic field in the CGM, e.g. as source of non-thermal pressure or by guiding anisotropic transport processes. As most observations can currently only probe the magnetic field out to distances of about $10\,\mathrm{kpc}$ from the disk for nearby edge-on spiral galaxies \citep[see, e.g.][]{Tuellmann2000,BeckR2012,Mao2012,Beck2015,DamasSegovia2016,Terral2017,Stein2019} there are very few observational constraints on the strength or the structure of the magnetic field in the CGM \citep{Bernet2008,Prochaska2019}. Moreover, cosmological simulations of galaxy formation that include magnetic fields and self-consistently reproduce the strength and structure of the magnetic field in the interstellar medium (ISM) have only recently become available \citep{Pakmor2014,Pakmor2017,Pakmor2018,Rieder2017,MartinAlvarez2018}.

Here, we use these existing simulations to analyse the strength and structure of the magnetic field in the CGM of Milky Way-like galaxies and their evolution with time. In Sec.~\ref{sec:auriga} we describe the suite of simulations we use in this paper with a focus on the modelling of  magnetic fields. In Sec.~\ref{sec:au6cgm} we look in detail at Au-6-CGM, a re-simulation of one of the \textsc{Auriga} galaxies with better spatial resolution in the CGM \citep{AurigaCGM}. In Sec.~\ref{sec:lvl3} we use the high-resolution sample of \textsc{Auriga} galaxies to understand the variation between different haloes. We show synthetic Faraday rotation maps of the CGM of Au-6-CGM in Sec.~\ref{sec:faraday} and compare them to observations. Finally we conclude in Sec.~\ref{sec:conclusion} with a summary and outlook.

\section{The \textsc{Auriga} simulations}
\label{sec:auriga}
The Auriga simulations are a suite of cosmological zoom-in simulations of Milky Way-like galaxies \citep{Auriga}. The zoom-in regions are centered around haloes with a halo mass between $10^{12}\,\mathrm{M_\odot}$ and $2\times10^{12}\,\mathrm{M_\odot}$ at $z=0$ selected from the dark matter only counterpart of the \textsc{Eagle} simulation box \citep{Schaye2005Eagle}. They were then re-simulated with a Lagrangian high resolution region around the haloes with an initial radius of about $1\,\mathrm{comoving\,Mpc/h}$. The closest low-resolution dark matter particle for the galaxies analysed in this paper is slightly more than $1\,\mathrm{Mpc}$ away at $z=0$. Here we use the set of six high resolution (level $3$) \textsc{Auriga} haloes that have a baryonic mass resolution of $6\times10^3\,\mathrm{M_\odot}$ and a dark matter mass resolution of $4\times10^{4}\,\mathrm{M_\odot}$ in the high resolution region. The different simulations are numbered as Au-X.

The \textsc{Auriga} simulations are run with the moving-mesh code \textsc{arepo} \citep{Arepo,Pakmor2016} that uses a second order finite volume scheme on an unstructured Voronoi mesh for gas and models dark matter, star particles, and black holes as collisionless particles. The \textsc{Auriga} model includes ideal magnetohydrodynamics (MHD) \citep{Pakmor2013,Pakmor2014} with the Powell 8-wave scheme for divergence control \citep{Powell1999}, self-gravity, primordial and metal-line cooling with self-shielding corrections \citep{Vogelsberger2013} and a time dependent spatially uniform UV background \citep{FaucherGiguere2009}, an effective subgrid model of the ISM for star formation and supernova feedback \citep{Springel2003}, an effective model for galactic winds \citep{Marinacci2014,Auriga}, chemical enrichment from AGB stars and from supernovae of type II and type Ia, and the formation and growth of supermassive black holes and their feedback as active galactic nuclei (AGN).

We focus first on the additional simulation Au-6-CGM \citep{AurigaCGM} from the \textsc{Surge} project that re-simulates Au-6 at the standard resolution of the \textsc{Auriga} project, but importantly enforces extra volume refinement in the CGM that guarantees that all cells have a volume smaller than $1\,\mathrm{kpc}^3$ out to a radius of $1.2\times R_{200,\mathrm{mean}}$, i.e. the radius within which the halo has an average density of $200$ times the mean density of the universe, which is equivalent to about $400\,\mathrm{kpc}$ at $z=0$. This increases the number of resolution elements in the CGM by more than an order of magnitude even compared to the high resolution \textsc{Auriga} simulations and has the additional benefit of essentially uniform spatial resolution in the CGM.

\begin{figure}
  \centering
  \includegraphics[width=\linewidth]{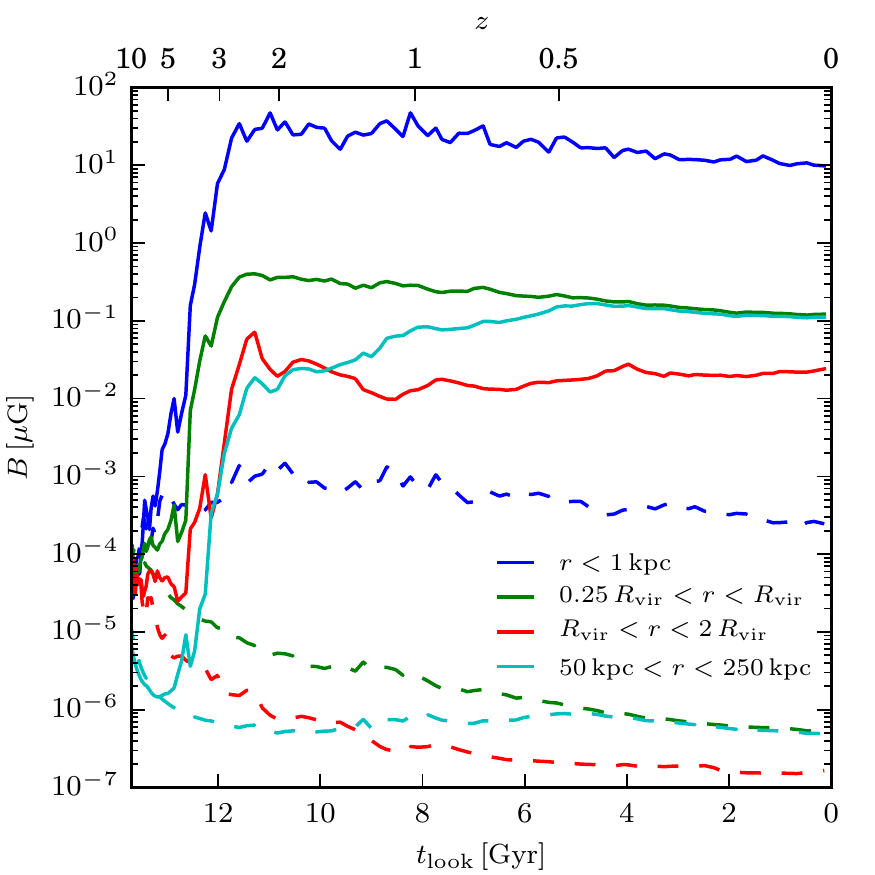}
  \caption{Time evolution of the volume averaged root mean square physical magnetic field strength (solid lines) for Au-6-CGM measured in a sphere with constant radius $1\,\mathrm{kpc}$ (blue line), in a spherical shell with $0.25\,R_\mathrm{vir}$ < $r$ < $R_\mathrm{vir}$ (green line), in a spherical shell with $R_\mathrm{vir}$ < $r$ < $2\,R_\mathrm{vir}$ (red line), and in a spherical shell with constant physical volume $50\mathrm{kpc}$ < $r$ < $250\mathrm{kpc}$ (cyan line). Dashed lines show the magnetic field strength in the same volumes if the magnetic field would change adiabatically, i.e. $B\propto \rho^{2/3}$. Note that the physical volume of the green and red lines changes over time, as $R_\mathrm{vir}$ expands. Satellite galaxies are excluded in the calculation.}
  \label{fig:amplification}
\end{figure} 

\begin{figure}
  \centering
  \includegraphics[width=\linewidth]{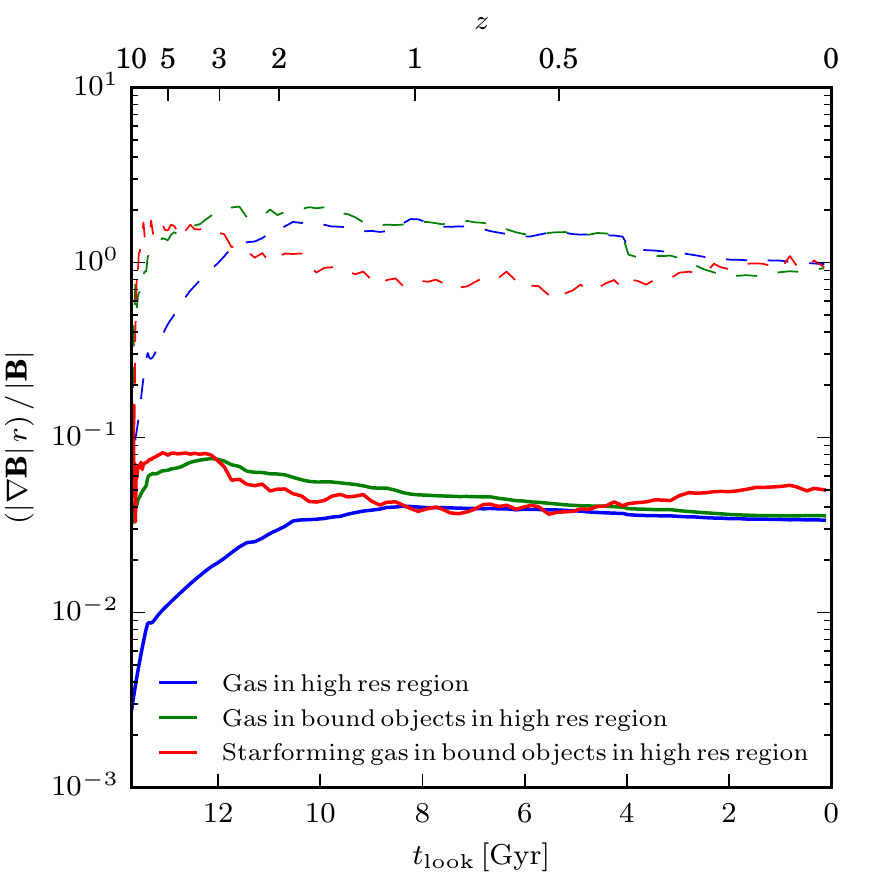}
  \caption{Time evolution of the median relative divergence error of the magnetic field (solid line) and its upper $3\sigma$ percentiles (dashed lines) for Au-6-CGM. The blue lines show all gas in the high resolution region as traced by a passive scalar. The green line shows all gas that in addition to being in the high resolution region is also part of bound objects as identified by subfind. Finally, the red line shows all gas that is in the high resolution region, in bound objects, and has a density larger than the star formation threshold $n_\mathrm{th}=0.1\,\mathrm{cm^{-3}}$.}
  \label{fig:berr_time}
\end{figure} 

\begin{figure}
  \centering
  \includegraphics[width=\linewidth]{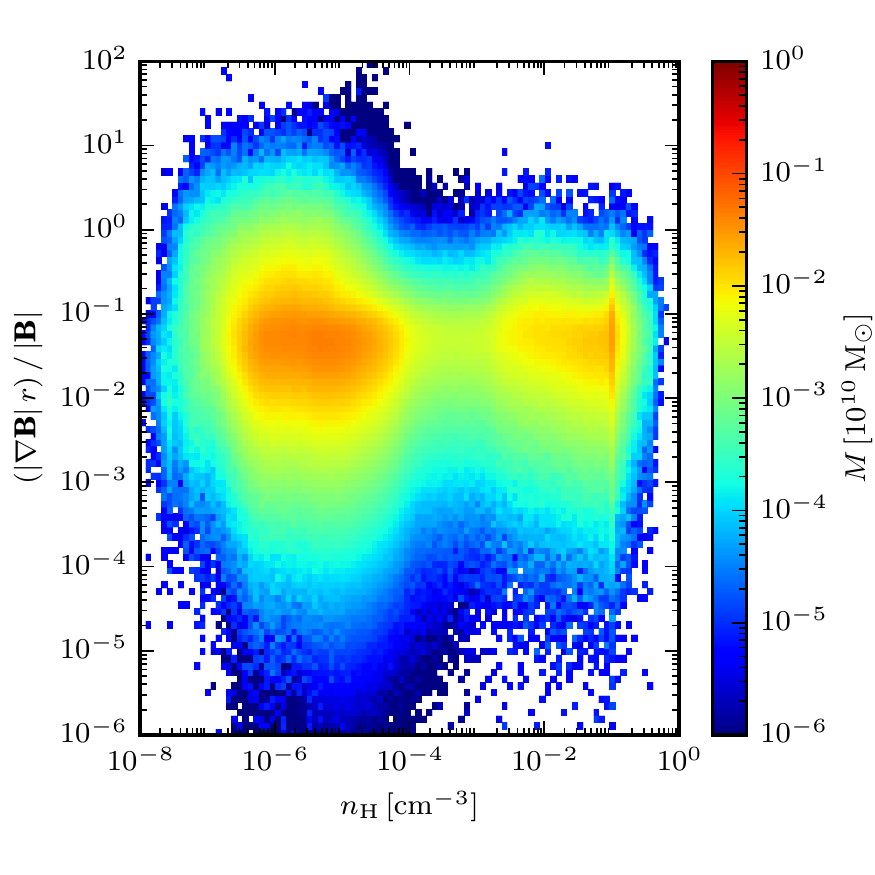}
  \caption{Mass weighted histogram of the relative divergence error of the magnetic field versus density for all gas cells in the high resolution region of Au-6-CGM at $z=0$. The typical error is of order of a few percent, consistent with the time evolution of the median relative divergence error shown in Fig.~\ref{fig:berr_time}, with a tail that extends above unity.}
  \label{fig:berr}
\end{figure} 

\begin{figure*}
  \centering
  \includegraphics[width=\linewidth]{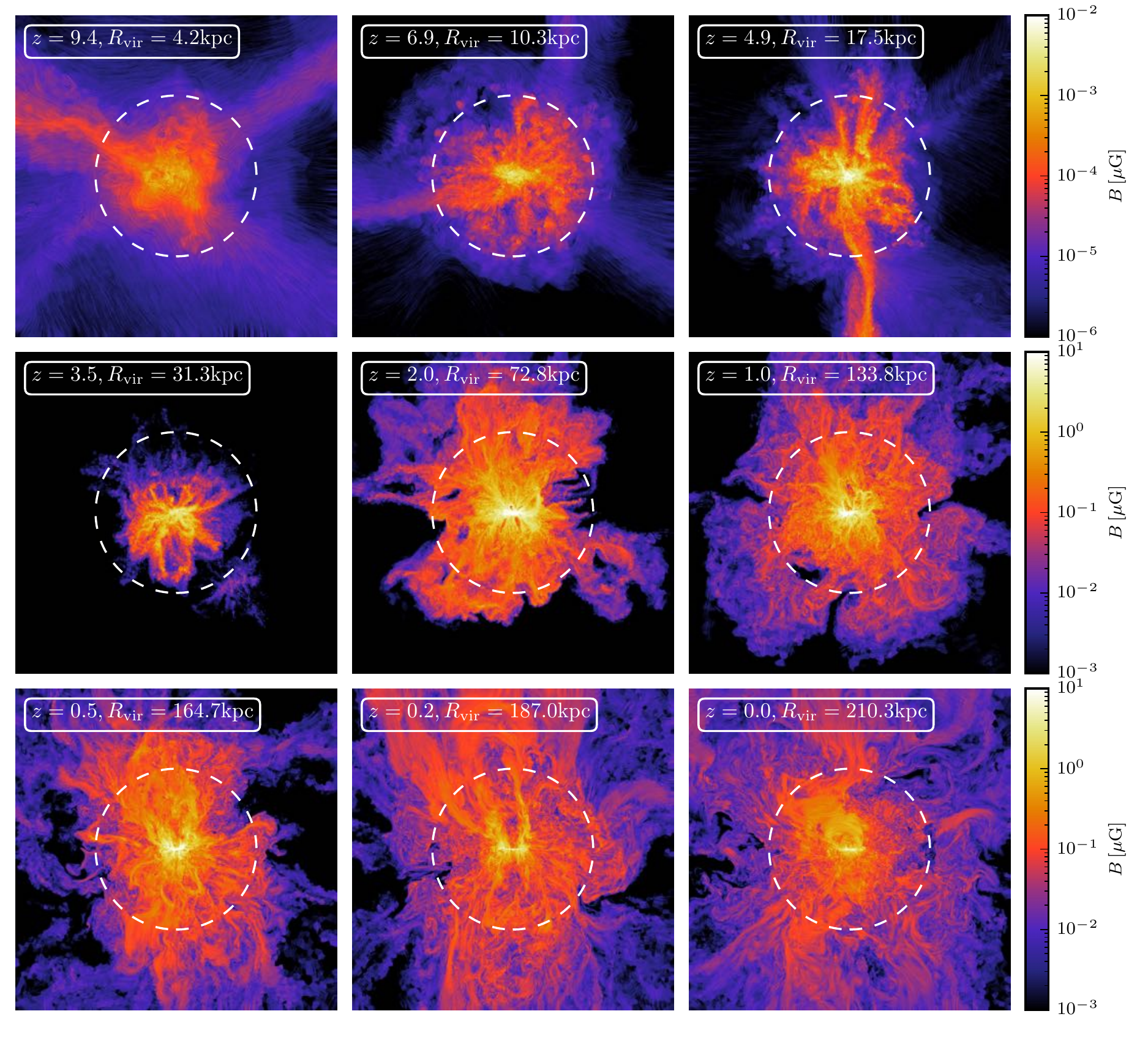}
  \caption{Thin edge-on projections of the volume average root mean square magnetic field strength for Au-6-CGM with a depth of $10\,\mathrm{kpc}$ at different redshifts. Superimposed on the magnetic field strength is the direction of the magnetic field using the line integral convolution method. The size of the projections is $4\,R_\mathrm{vir}$ at each redshift and the virial radius is shown by the dashed circle. Note that the top row and the two lower rows use different ranges of $B$ values.}
  \label{fig:proj_au6_bfld}
\end{figure*}

\begin{figure*}
  \centering
  \includegraphics[width=\linewidth]{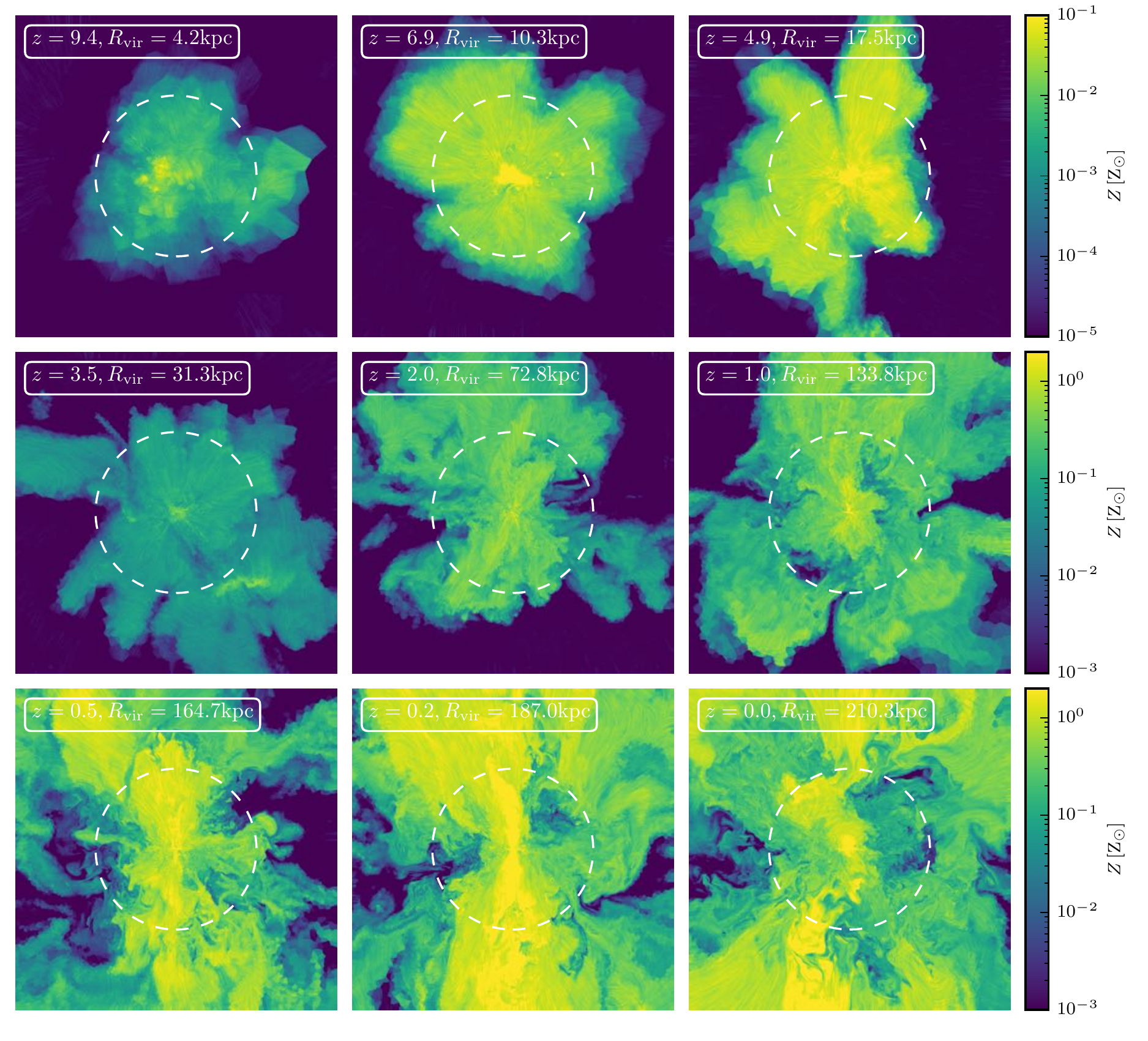}
  \caption{Thin edge-on projections of the mass weighted average metallicity for Au-6-CGM with a depth of $10\,\mathrm{kpc}$ at different redshifts. Superimposed on the metallicity is the direction of the velocity field using the line integral convolution method. The size of the projections is $4\,R_\mathrm{vir}$ at each redshift and the virial radius is shown by the dashed circle. Note that the top row and the two lower rows use different ranges of $Z$ values.}
  \label{fig:proj_au6_metal}
\end{figure*}

\begin{figure*}
  \centering
  \includegraphics[width=\linewidth]{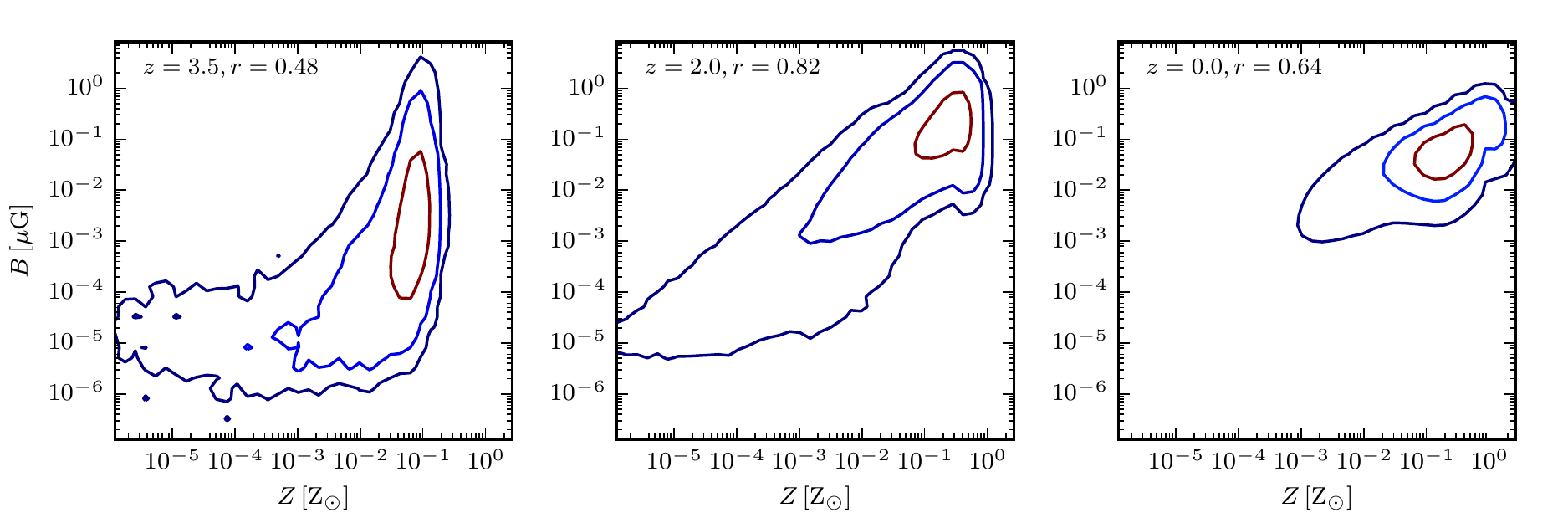}
  \caption{Contours of the volume weighted histogram of magnetic field strength and metallicity for Au-6-CGM for $0.25 R_\mathrm{vir}$ < $r$ < $R_\mathrm{vir}$. The contours show the area containing $50\%$ (brown), $90\%$ (light blue), and $99\%$ (dark blue) of the volume. The label in the top left corner shows the redshift $z$ and the Pearson coefficient $r$ of the correlation between logarithmic magnetic field strength and logarithmic metallicity.}
  \label{fig:bfldmetals}
\end{figure*}

Here, we are particularly interested in the evolution of magnetic fields in the CGM around the main galaxy. The magnetic fields are seeded at the start of the simulation at $z=127$ with a uniform seed field with a comoving strength of $10^{-14}\,\mathrm{G}$ that is equivalent to a physical strength of $1.6\times10^{-10}\,\mathrm{G}$. The strength of the seed field is chosen to be large enough that the turbulent dynamo in the galaxy saturates well before $z=2$ when the disk forms \citep{Pakmor2014} and small enough that the seed field itself does not affect the galaxy \citep{Marinacci2016}.

The simulation self-consistently follows the ideal MHD equations for their evolution. After the galaxy forms, a turbulent dynamo driven by accretion of gas on to the new galaxy sets in at $z\leq5$ and amplifies the magnetic field until it saturates at a magnetic energy of roughly $10\%$ of the turbulent kinetic energy \citep{Pakmor2017}. For the high resolution simulations this process starts between $z=10$ and $z=7$ and saturation of the magnetic energy density at about $10\%$ of the turbulent energy density is reached in the center of the galaxy before $z=3$, consistent with analytical models of a turbulent dynamo in high redshift galaxies \citep{Schober2013}. The magnetic field in the ISM is then further amplified and ordered after the galaxy forms a gas disk. This process starts between $z=2$ and $z=1$ and saturates the magnetic field strength roughly in equipartition with the turbulent energy density around $z=0.5$ for most of the disk.

Complementary to previous work that focused on the evolution of the magnetic field in the ISM of the galaxies \citep{Pakmor2017}, here we focus on the evolution of the magnetic field in the CGM. We loosely define the CGM as the gas around the galaxy that is directly influenced by it, i.e. the gas out to the farthest distance galactic winds can reach. The most important process here is the galactic wind, driven by ongoing star formation and possibly AGN activity, that drives gas out of the galaxy into the CGM. We employ an effective model for galactic winds in which star formation leads to the formation of wind particles, modelling the energy input from massive stars and core collapse supernovae. The wind particles are imparted with momentum and move in a random direction until they recouple and deposit their momentum (and some amount of thermal energy and metals) once the cells they currently reside in have a density smaller than $5\%$ of the threshold density for star formation ($n_\mathrm{th}=0.1\,\mathrm{cm^{-3}}$) \citep{Marinacci2014,Auriga}. At $z=0$ the recoupling of the wind particles happens at a height of $\left| z \right| \lesssim 10\,\mathrm{kpc}$ above and below the disk. The galactic wind driven by the wind particles is the main source of outflows from the galaxy for most of its evolution (AGN are generally subdominant at this halo mass in the \textsc{Auriga} simulations possibly as a result of the smooth ISM model.). At low redshift the galactic wind becomes mostly bipolar, which is not directly set by the isotropic wind model, but an emergent phenomenon as the wind takes the path of least resistance away from the galaxy.

Note that the wind particles do not transport any magnetic energy, so the magnetisation of the galactic wind has its origin in the magnetisation of the gas to which the wind particles recouple. This gas, by construction of the wind model, is not the starforming ISM but has a density lower than the star formation threshold. More realistic galactic winds in future simulations could instead be able to transport gas directly from the ISM that has a different magnetisation than the lower density gas surrounding it.
 
\section{A case study: Au-6-CGM}
\label{sec:au6cgm}

We first concentrate on Au-6-CGM with its exquisite spatial resolution of $1\,\mathrm{kpc}$ (i.e. cells with a physical volume of $1\,\mathrm{kpc^3}$) in the CGM out to a distance of $400\,\mathrm{kpc}$ at $z=0$ \citep{AurigaCGM}. The average projected magnetic field strength of Au-6-CGM at z=0 is shown in an edge-on projection in Fig.~\ref{fig:cgm} (left panel). It clearly shows that the CGM is highly magnetised at this time, i.e. its strength is several orders of magnitudes stronger than the field expected for pure adiabatic contraction of the seed field. It also suggests that outflows transport highly magnetised gas from near the disk into the CGM well beyond the virial radius of the galaxy ($R_\mathrm{vir} = R_{200,\mathrm{crit}}$, i.e. the radius within which the halo has an average density of $200$ times the critical density of the universe). Moreover, within the virial radius the magnetic field strength is mostly homogenous with azimuth, unlike at larger distances where it varies significantly. Within the virial radius the magnetic field appears to be mostly turbulent. Only in coherent large-scale outflows does it become ordered along the direction of the flow as can be seen above the disk.

It is useful to compare the magnetisation of the gas to its metallicity, as the latter rather directly traces outflows from the galaxy and cannot be produced in the halo itself. Comparing the magnetic field strength to the metallicity, which is shown in the right panel of Fig.~\ref{fig:cgm}, we can see that at distances larger than the virial radius the magnetisation of the gas correlates with its metal enrichment. Within the virial radius they seem to be mostly decoupled at $z=0$, as the metallicity of the gas is much higher in outflows than in the rest of the halo. Note that the average magnetic field strength is computed as the constant magnetic field strength that has the same total magnetic energy in the column of a pixel and the average metallicity is computed as the constant metallicity that has the same total metal mass in the column of a pixel as the actual simulation.

In addition to the qualitative analysis, however, it is necessary to analyse the magnetic field in the CGM quantitatively to understand in detail how the CGM becomes magnetised and how its magnetic field changes over time.

\begin{figure}
  \centering
  \includegraphics[width=\linewidth]{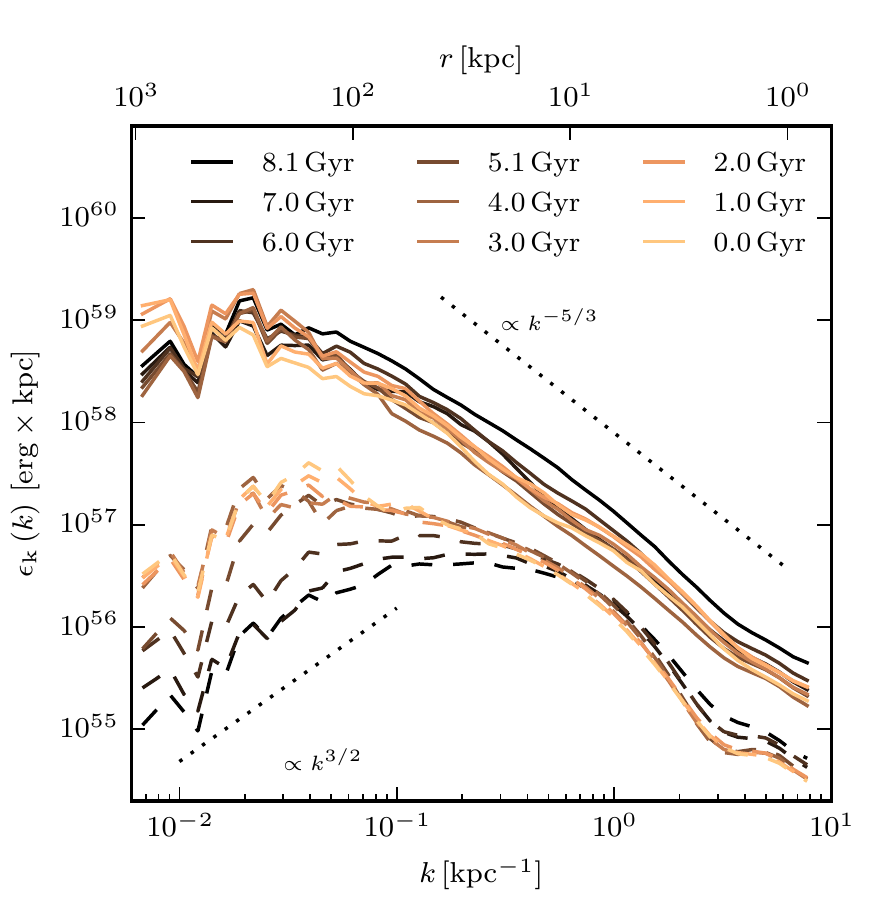}
  \caption{Kinetic (solid lines) and magnetic (dashed lines) energy power spectra of Au-6-CGM at different times computed in a fixed physical volume of a spherical shell with $50\,\mathrm{kpc}$ < $r$ < $250\,\mathrm{kpc}$. The black dotted lines show the slopes of a Kolmogorov spectrum ($\propto k^{-5/3}$) \citep{Kolmogorov1941} and a Kazantsev spectrum ($\propto k^{3/2}$) \citep{Kazantsev1985} that are theoretically expected for a subsonic turbulent dynamo.}
  \label{fig:powerspectra}
\end{figure}

\begin{figure*}
  \centering
  \includegraphics[width=\linewidth]{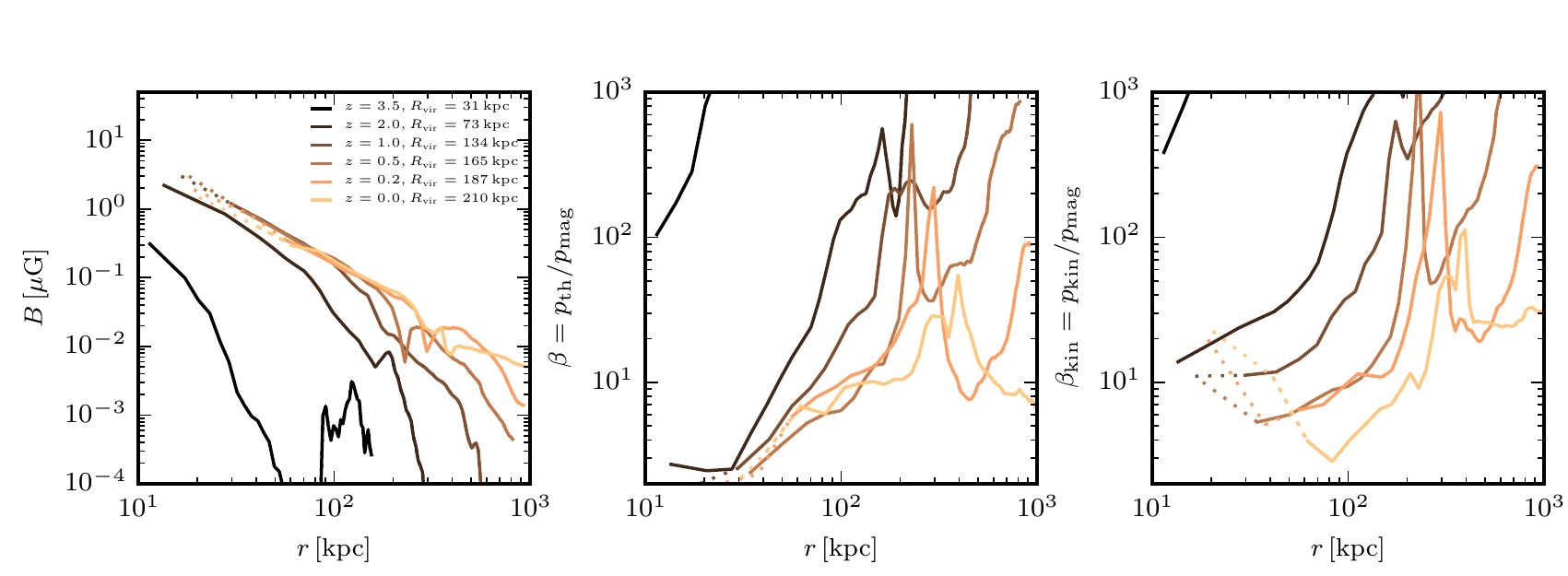}
  \caption{Time evolution of the radial profiles of Au-6-CGM of the magnetic field strength (left panel) and ratios of thermal pressure over magnetic pressure (middle panel) and kinetic pressure over magnetic pressure (right panel) at different redshifts. The kinetic pressure has been computed as $p_\mathrm{kin} = \frac{1}{2V} \int \rho \mathbfit{v}^2 dV$ in the rest frame of the galaxy. The dotted lines indicate distances smaller than $0.25\,R_\mathrm{vir}$ which roughly coincides with the radius of the gas disk. Satellite galaxies have been excluded from the profiles.}
  \label{fig:profilestime}
\end{figure*}

\subsection{Amplification of the magnetic field strength in the CGM}

The evolution of the magnetic field strength in the center of Au-6-CGM, in the halo within $R_\mathrm{vir}$, and around the halo outside $R_\mathrm{vir}$ is shown in Fig.~\ref{fig:amplification}. The magnetic field is efficiently amplified first in the center of the galaxy, as discussed in detail in \citet{Pakmor2017}. Once the center of the galaxy has been magnetised, the magnetic field strength in the halo and the region around increases as well. The average magnetic field strength around the virial radius of the halo does not change much after $z\approx 2$ at a typical average strength of a few $10^{-2}\,\mathrm{\mu G}$. This strength is about a thousand times weaker than the magnetic field strength in the center of the halo \citep{Pakmor2017}.

There are two obvious but fundamentally different mechanisms that are, in principle, able to magnetise the CGM. Outflows that efficiently transport gas that originates from the ISM into the CGM become magnetised once the turbulent dynamo has amplified the magnetic field in the ISM. They mix highly magnetised gas into the CGM, thereby magnetising it. This mode of magnetising the CGM by magnetised ouflows from the galaxy is always present because the magnetic field in the ISM and in the gas surrounding the starforming ISM that is picked up by the galactic wind is larger than in the CGM. In addition, the gas in the CGM can be highly turbulent. Thus, there may also be an in-situ dynamo at work that amplifies an existing magnetic field in the CGM. To understand if outflows are sufficient to explain the magnetisation of the CGM or a dynamo is acting in the halo in addition, we need to look in more detail at the spatial distribution of the magnetic field strength and the structure of the magnetic field in the CGM.

For diagnostic purposes we show the median and upper $3\sigma$ percentile of the time evolution of the relative divergence error of the magnetic field in Fig.~\ref{fig:berr_time} and its correlation with density at $z=0$ in  Fig.~\ref{fig:berr}. The typical divergence error is of the order of a few percent, independent of density and almost independent of time. The $3\sigma$ or $0.15\%$ largest values are of order unity, again essentially independent of time. At low densities ($n_\mathrm{H}<10^{-4}\mathrm{cm^{-3}}$) the maximum error can become larger as gradients towards the low resolution region become poorly resolved, similar to the isolated galaxies in \citet{Pakmor2013} where the relative divergence error is also clearly larger at the edge of the gas disk where gradients are poorly resolved. Provided that the tail of the distribution does not play a role in driving the dynamo, we conclude that our results should be unaffected by the numerical divergence of the magnetic field.

\subsection{The spatial extent of the magnetised CGM}

The time evolution of the magnetic field in the CGM of Au-6-CGM is shown as thin projections in Fig.~\ref{fig:proj_au6_bfld}. For comparison we also show the time evolution of the metallicity in the CGM in Fig.~\ref{fig:proj_au6_metal}. At very high redshift ($z > 7$) the magnetic field only changes adiabatically as gas expands or is compressed. The structure of the magnetic field is still strongly influenced by the uniform initial seed field and only changed by an already turbulent velocity field in the young halo. After $z=7$ the center of the galaxy becomes well enough resolved for a turbulent dynamo to operate and the magnetic field strength quickly amplifies and saturates (before $z=3$) in the center \citep{Pakmor2017}. Outflows driven by star formation that originate from the center of the galaxy then push magnetised gas out into the CGM and slightly beyond the virial radius. The magnetisation of the outflows increases as the magnetic field strength in the galaxy increases. This can be seen comparing the magnetic field strength in the outflows, for example, at $z=5$ and $z=2$ in Fig.~\ref{fig:proj_au6_bfld}. At the same time these outflows are already enriched with metals, thus metal enrichment and magnetisation of the CGM by mixing of outflows with CGM gas go hand in hand. Consequently the structures seen in magnetisation and metallicity correlate strongly. We conclude that, before $z\approx1$, the magnetisation of the CGM is dominantly driven by outflows from the galaxy.

At $z=1$ outflows have pushed magnetised and metal enriched gas out to distances of $300\,\mathrm{kpc}$, larger than $2\,R_\mathrm{vir}$, though the magnetisation at large distances remains patchy. At the same time the magnetic field strength within $R_\mathrm{vir}$ has become essentially uniform, even though outflows are still visibly more metal enriched in the same area than the background gas. This may be a direct contribution of a turbulent dynamo operating in the halo, which sets the field strength within the halo. Beyond the virial radius the magnetic field strongly correlates with metallicity. There are still regions with primordial gas, i.e. not significantly enriched with metals and not magnetised beyond the seed field, just beyond the virial radius of the halo at $z=1$.

At lower redshift this changes as essentially all the volume around the halo out to at least $2\,R_\mathrm{vir}$ becomes magnetised. Until $z=0.5$ the magnetic field in the CGM remains completely unordered as there are no large scale structures in the velocity field that could drive an ordering of the magnetic field. The magnetic field in these outflows only becomes ordered along the direction of the outflow at $z\approx 0.5$, when the galactic disk of Au-6-CGM develops large-scale coherent outflows that are stable for many Gyrs.

We attempt to quantify the relation between metallicity as a tracer of outflows and magnetic field strength in Fig.~\ref{fig:bfldmetals}. At $z=3.5$ all the gas in the CGM, which already has an amplified magnetic field strength, is also highly metal enriched. This strongly argues that the magnetised gas in the CGM at this time had its magnetic field amplified in the ISM and was then ejected by outflows. At $z=2$ metallicity and magnetic field strength are well correlated over many orders of magnitude in the CGM. The Pearson correlation coefficient $r$ between the logarithmic magnetic field strength and the logarithmic metallicity has increased from $0.48$ at $z=3.5$ to $0.82$ at $z=2$. In contrast, at $z=0$ magnetic field strength and metallicity in the CGM are less strongly correlated and the correlation coefficient has dropped to $0.64$. Interestingly the correlation between magnetic field strength and metallicity at $z=2$ looks very similar to the correlation found for an isolated disk galaxy by \citet{Butsky2017}.

\subsection{The amplification mechanism}

To understand whether an in-situ turbulent dynamo is really operating in the CGM of Au-6-CGM at low redshift, we analyse power spectra of magnetic and kinetic energy in Fig.~\ref{fig:powerspectra} that are computed from a spherical shell with a constant physical extent of $50\,\mathrm{kpc}$ < $r$ < $250\,\mathrm{kpc}$ in a zero-padded box with a size of $1\,\mathrm{Mpc}$ by taking the absolute square of the Fourier transforms of $\mathbfit{B}/\sqrt{8 \pi}$ and $\sqrt{\rho} \mathbfit{v}$, respectively. We checked that including the central $50\,\mathrm{kpc}$ does not qualitatively change our results. The kinetic power spectra clearly show that the gas in the CGM is turbulent. Turbulence is driven on scales of several $100\,\mathrm{kpc}$, likely by a combination of strong outflows from the disk \citep{Fielding2016} as well as inflowing gas from the cosmic web on to the halo \citep{Klessen2010,Iapichino2013} and potentially torques from massive satellite galaxies as has been argued for in galaxy clusters \citep{Kim2007,Ruszkowski2011,Miniati2015}. The power spectrum follows the expected slope for subsonic Kolmogorov turbulence down to the resolution limit at $\approx 1\,\mathrm{kpc}$. At a look-back time of $8\,\mathrm{Gyr}$ the turbulence is already fully established in the CGM and changes very little down to $z=0$. 

The magnetic energy, in contrast, shows clear signs of an ongoing turbulent dynamo that amplifies the magnetic field strength. The amplification of the magnetic field in the halo is directly visible in Fig.~\ref{fig:amplification} in the difference between between the actual field strength and field strength expected for the adiabatic evolution of the magnetic field. Fig.~\ref{fig:amplification} also indicates that the in-situ dynamo sets in already around $z=2$. Fig.~\ref{fig:powerspectra} shows that the magnetic energy density is already saturated on small scales at a look-back time of $8\,\mathrm{Gyr}$. As discussed above, the field that is already saturated on small scales is a result of outflows of magnetised gas that carry the magnetic field that has been amplified in the center of the galaxy by a fast turbulent dynamo \citep{Pakmor2017} and then spreads out until the magnetic field is picked up by the galactic wind.

In a second step the dynamo in the halo that starts with an already saturated magnetic field on small scales pushes magnetic energy to larger scales. On large scales the magnetic field is consistent with the Kazantsev spectrum \citep{Kazantsev1985} expected for a turbulent dynamo. Its strength increases linearly with time until it saturates at a look-back time of about $4\,\mathrm{Gyr}$.

At saturation, the magnetic energy is about $10\%$ of the turbulent kinetic energy on scales smaller than the peak of the magnetic power spectrum, typical for a subsonic turbulent dynamo \citep{Federrath2016}, and about $1\%$ of the kinetic energy at scales larger than the injection scale of kinetic energy. We therefore conclude that at late times the magnetic field strength in the CGM within $R_\mathrm{vir}$ is set by a halo-wide turbulent dynamo in the linear phase that operates on timescales of Gigayears.

\begin{figure}
  \centering
  \includegraphics[width=\linewidth]{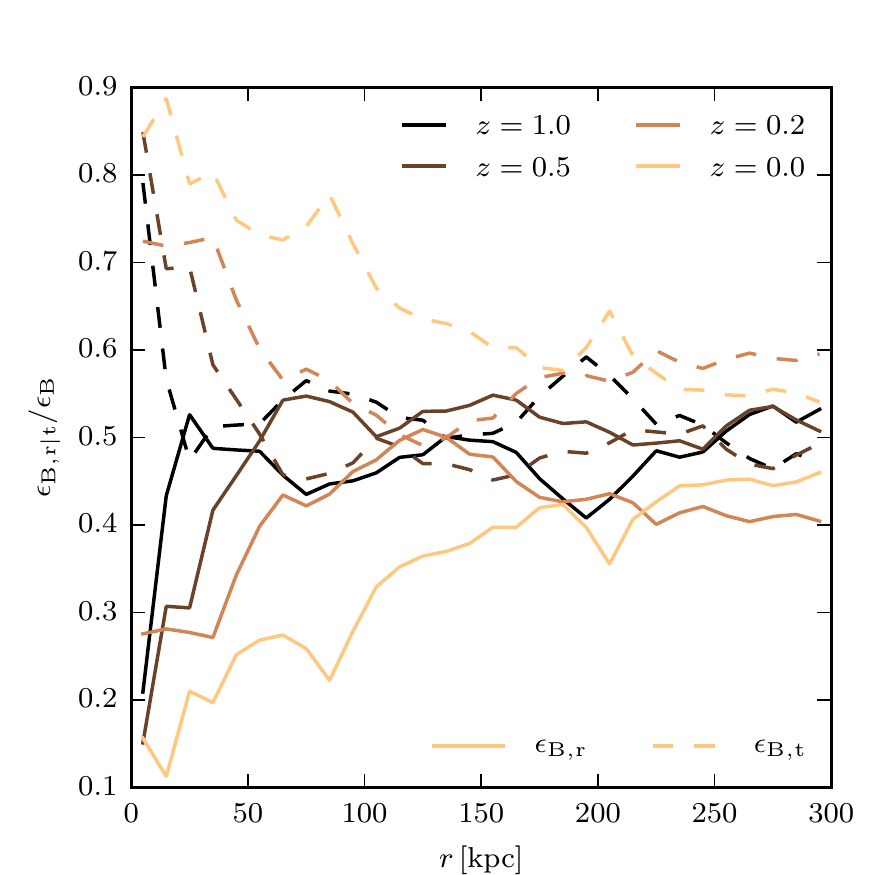}
  \includegraphics[width=\linewidth]{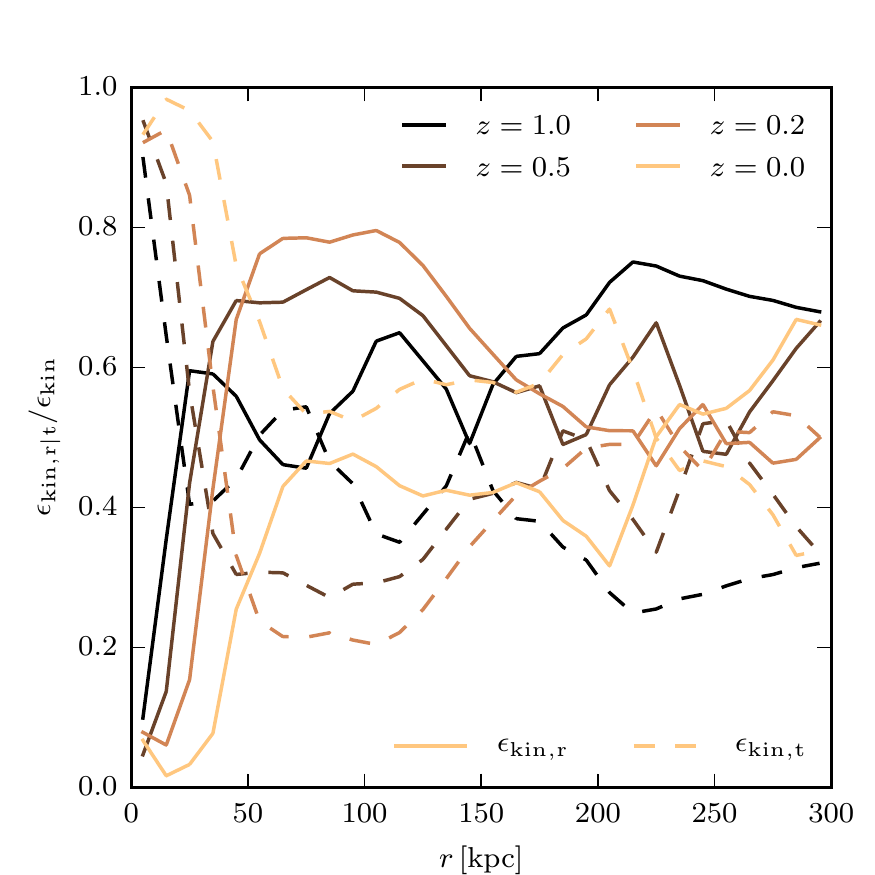}
  \caption{Time evolution of the radial profile of the fraction of the radial and tangential components of the total magnetic (top panel) and total kinetic energy (bottom panel) in spherical shells for Au-6-CGM. Here the radial component is computed in spherical coordinates and the tangential component is defined as the full magnetic field vector minus the radial component.}
  \label{fig:radial}
\end{figure}

The evolution of the radial profile of the magnetic field strength and the ratios between magnetic pressure and thermal and kinetic pressure, respectively, are shown in Fig.~\ref{fig:profilestime}. As the halo grows, the magnetic field strength in the CGM first grows at fixed physical radius until $z=1$, after which it remains essentially constant in the inner part of the CGM, but keeps growing in the outer parts. Meanwhile, the radius out to which the galaxy magnetises its environment, as marked by a steep decrease of the magnetic field strength, continues to grow until $z=0$ when it has reached a distance of more than $500\,\mathrm{kpc}$. At $z=2$ the magnetic pressure reaches about $10\%$ of the thermal and kinetic pressure, i.e. $\beta \approx 10$ in the CGM at the virial radius of then $70\,\mathrm{kpc}$ with smaller values at smaller radii and larger $\beta$ at larger distances. At low redshift ($z < 0.2$) the magnetic pressure reaches $\beta \approx 10$ in most of the CGM from $R\approx 50\,\mathrm{kpc}$ out to $2\,R_\mathrm{vir}$.

Thus, in the CGM the magnetic energy density is at typically an order of magnitude below equipartition with thermal or turbulent energy density, but more important close to the disk. Nevertheless, it is large enough so that it cannot be completely ignored \citep[see, e.g.][]{Berlok2019}. The main difference to the ISM where the magnetic field reaches equipartition \citep{Pakmor2017} is the lack of any large-scale galactic dynamo that can amplify the magnetic field beyond the saturation strength of the turbulent dynamo, which always saturates significantly below equipartition \citep{Federrath2016}.

\subsection{The orientation of the magnetic field in the CGM}

As seen most easily in Fig.~\ref{fig:proj_au6_metal} at late times the galaxy generates strong, coherent, mostly bipolar outflows. To quantify how this changes the preferred orientation of the magnetic field and velocity field in the CGM, we show radial profiles of the average ratios of the magnetic and kinetic energy of the radial and tangential component in Fig.~\ref{fig:radial}. Here, we would see a radial energy fraction of $1/3$ for an isotropic field. Owing to the disk, the kinetic energy (bottom panel) is completely dominated by the tangential component at small radii $r < 50\,\mathrm{kpc}$. At larger radii, in the CGM, the kinetic energy has a significant preference for radial motions, primarily caused by outflows from the disk that are denser and faster than the background CGM.

\begin{figure}
  \centering
  \includegraphics[width=\linewidth]{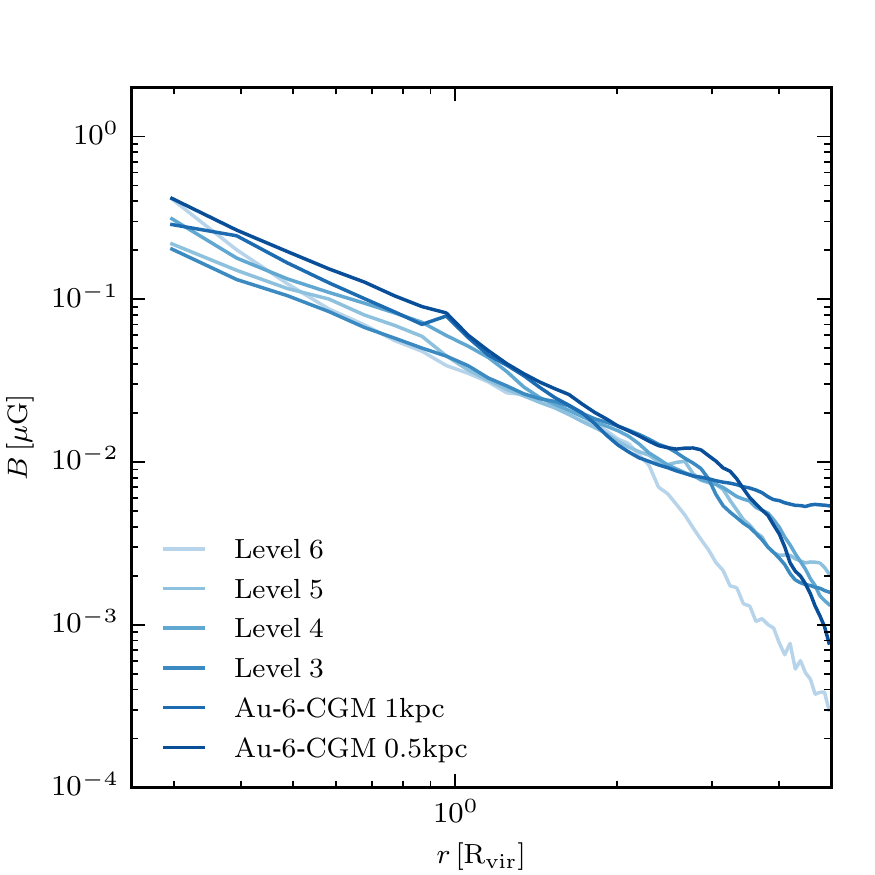}
  \caption{Radial profiles of the magnetic field strength for Au-6 at different resolutions from level 6 ($M_\mathrm{baryons} = 3\times10^6\,\mathrm{M_\odot}$) to level 3 ($M_\mathrm{baryons} = 6\times10^3\,\mathrm{M_\odot}$) $z=0$ in units of the virial radius. Also included are radial profiles of Au-6-CGM ($V_\mathrm{cell} \leq 1\,\mathrm{kpc}^3$ out to $400\,\mathrm{kpc}$) and an additional simulation with $V_\mathrm{cell} \leq \left( 0.5\,\mathrm{kpc} \right)^3$ for which this high resolution CGM region only reaches out to $R_\mathrm{vir} \approx 200\,\mathrm{kpc}$ at $z=0$. Satellite galaxies have been excluded from the profiles.}
  \label{fig:resolution}
\end{figure}

\begin{figure}
  \centering
  \includegraphics[width=\linewidth]{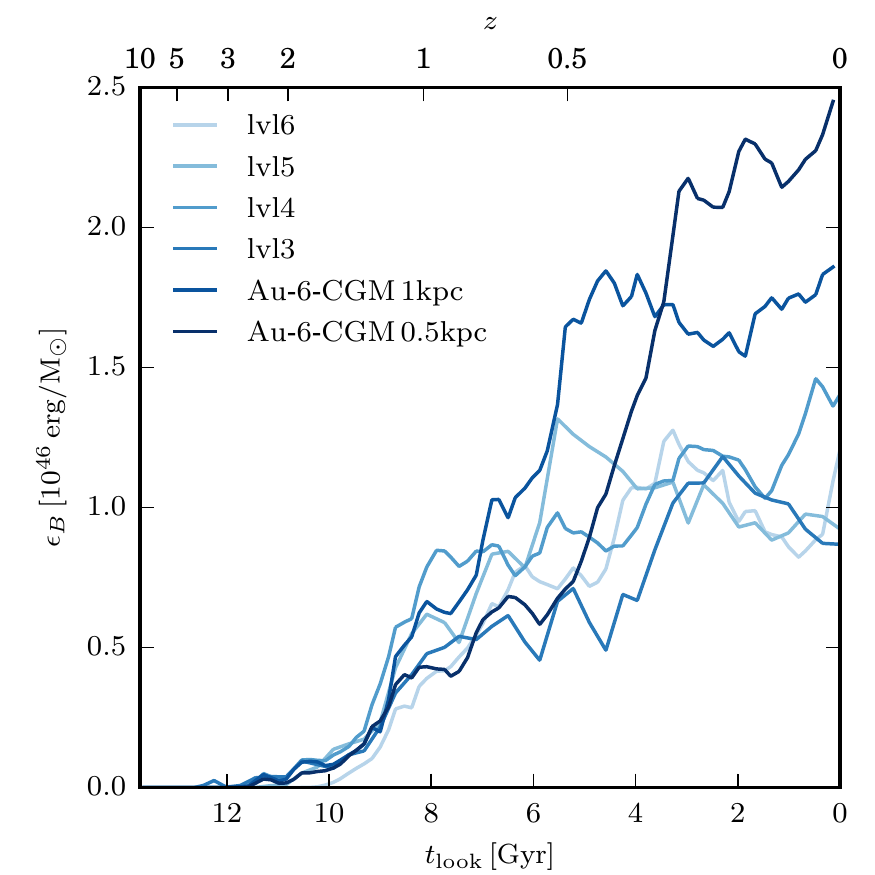}
  \caption{Time evolution of the total specific magnetic energy (total magnetic energy divided by total gas mass) in a spherical shell with constant physical extent $50\,\mathrm{kpc} < r < 250\,\mathrm{kpc}$. We show the evolution for Au-6 at different resolutions from level 6 ($M_\mathrm{baryons} = 3\times10^6\,\mathrm{M_\odot}$) to level 3 ($M_\mathrm{baryons} = 6\times10^3\,\mathrm{M_\odot}$) as well as for Au-6-CGM ($V_\mathrm{cell} \leq 1\,\mathrm{kpc}^3$ out to $400\,\mathrm{kpc}$) and an additional simulation with $V_\mathrm{cell} \leq \left( 0.5\,\mathrm{kpc} \right)^3$ for which this high resolution CGM region only reaches out to $R_\mathrm{vir} \approx 200\,\mathrm{kpc}$ at $z=0$.}
  \label{fig:resolution_amplification}
\end{figure}

There is also more energy in the radial component of the magnetic field  (top panel) than expected for a completely isotropic field, though the magnetic field is closer to isotropy than the velocity field. Moreover, even when the kinetic energy is dominated by its radial component at certain radii the radial magnetic field there is not much stronger than at other radii. This is consistent with our previous finding that the magnetic field strength is much more homogeneous within the virial radius of the halo (see also Fig.~\ref{fig:cones}) and the magnetic field strength in the outflows is only slightly enhanced over the background, owing to the in-situ dynamo in the halo. 

\subsection{Convergence}

Since many processes in numerical simulations and in particular the numerical modelling of turbulence are affected by resolution we show in Fig.~\ref{fig:resolution} the radial profile of the magnetic field strength in the CGM of Au-6 (without additional spatial refinement) at $z=0$ for different resolution levels and including Au-6-CGM with two different minimum spatial resolutions of $1\,\mathrm{kpc}$ and $0.5\,\mathrm{kpc}$. The normalisation and the slope of all profiles are very similar out to $2\,R_\mathrm{vir}$. For the standard \textsc{Auriga} simulations with a purely Lagrangian refinement criterion, i.e. constant mass per cell in the high resolution zoom-in region, the magnetic field strength varies by about a factor of two between the simulations with different resolution, but without an obvious trend with resolution. The two simulations Au-6-CGM with additional refinement in the CGM that enforce a minimum spatial resolution are at the high magnetic field strength end of these variations, though they show a very similar slope of the profile as the standard \textsc{Auriga} simulations. 

The time evolution of the total specific magnetic energy in a constant physical volume $50\,\mathrm{kpc} < r < 250\,\mathrm{kpc}$ for the same runs is shown in Fig.~\ref{fig:resolution_amplification}. Consistent with Fig.~\ref{fig:amplification} they all show a linear increase with time starting around $z=2$ and saturating around or shortly after $z=0.5$. As discussed above, we argue that this linear increase is a signature of an in-situ turbulent dynamo in the halo that is already saturated on the smallest scales. Similar to the radial profiles at $z=0$ the time evolution of the magnetic fields strength for the standard \textsc{Auriga} simulations is very similar (lvl2-lvl6), without any obvious trend with resolution. This is consistent with the interpretation that a dynamo amplifies the magnetic field that is seeded by the galactic wind linear in time. By construction, the substantial magnetic seed field in the CGM is well resolved and saturated on the smallest resolved scales. Its further evolution is then resolved by construction. Interestingly the two simulations with additional CGM refinement saturate at a slightly but significantly larger specific magnetic energy.

\begin{figure}
  \centering
  \includegraphics[width=\linewidth]{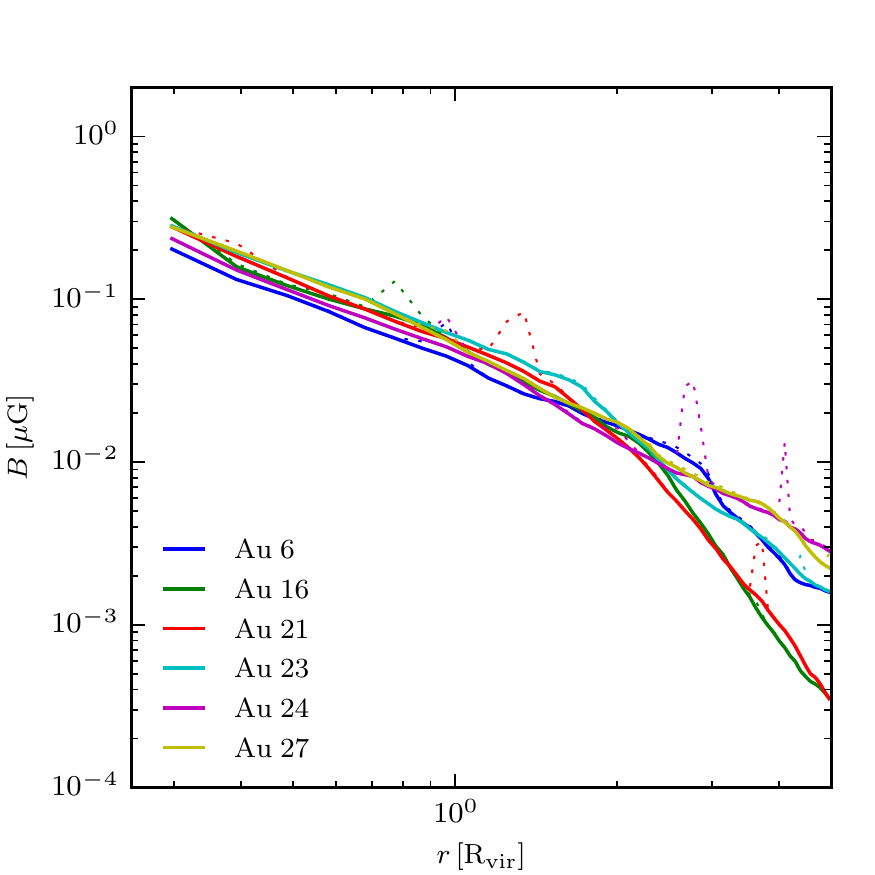}
  \includegraphics[width=\linewidth]{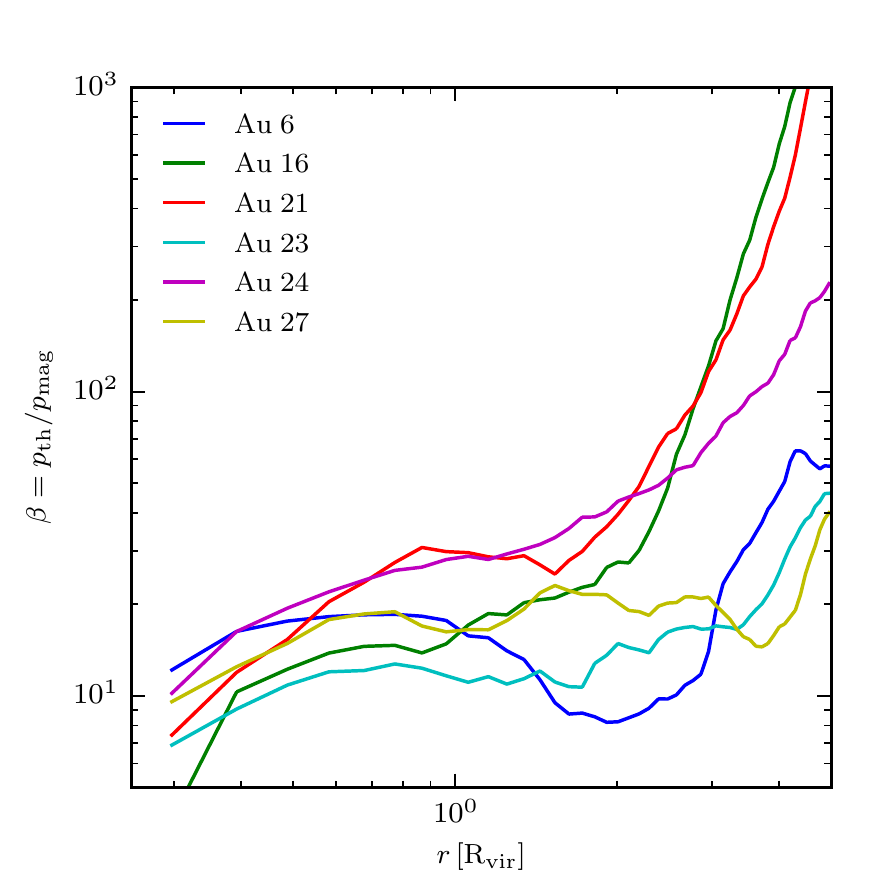}
  \caption{Radial profiles of the magnetic field strength (top panel) and plasma beta (bottom panel) for the $6$ high resolution \textsc{Auriga} haloes at $z=0$ in units of the virial radius. The solid lines show the profiles excluding any satellite galaxies in the main halo as well as any other galaxies. The thin dotted lines in the top panel show the radial profile of the magnetic field strength for all cells instead. The profiles are cut at $0.25\,R_\mathrm{vir}$ to exclude the central galaxy.}
  \label{fig:profiles}
\end{figure}

\begin{figure}
  \centering
  \includegraphics[width=\linewidth]{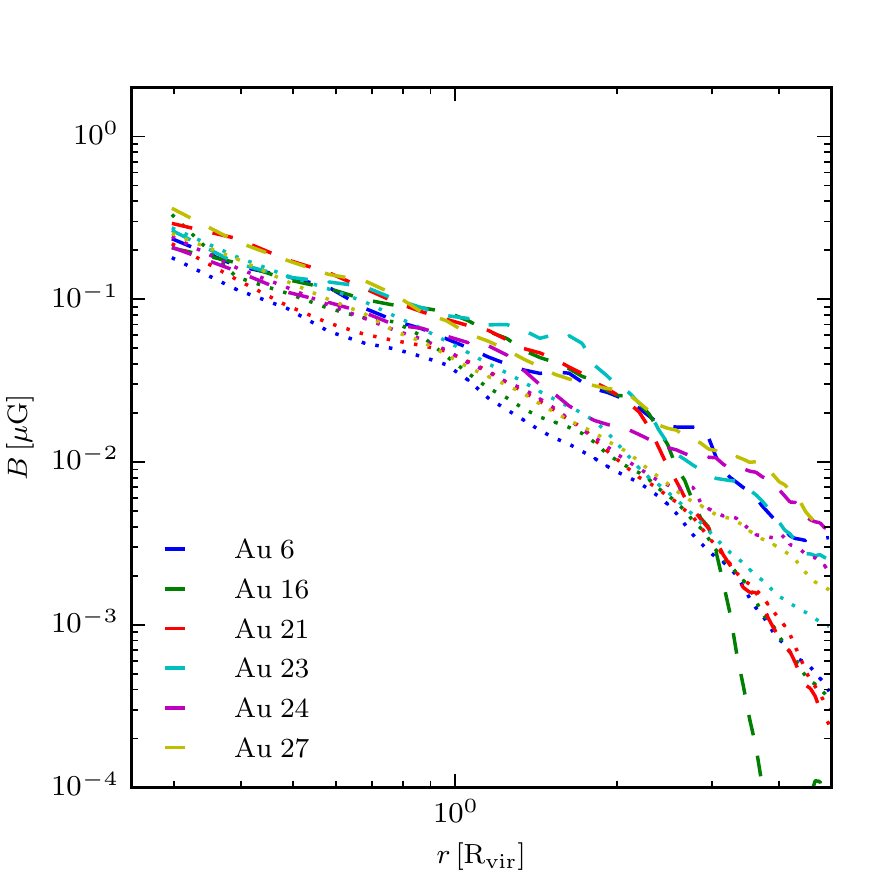}
  \includegraphics[width=\linewidth]{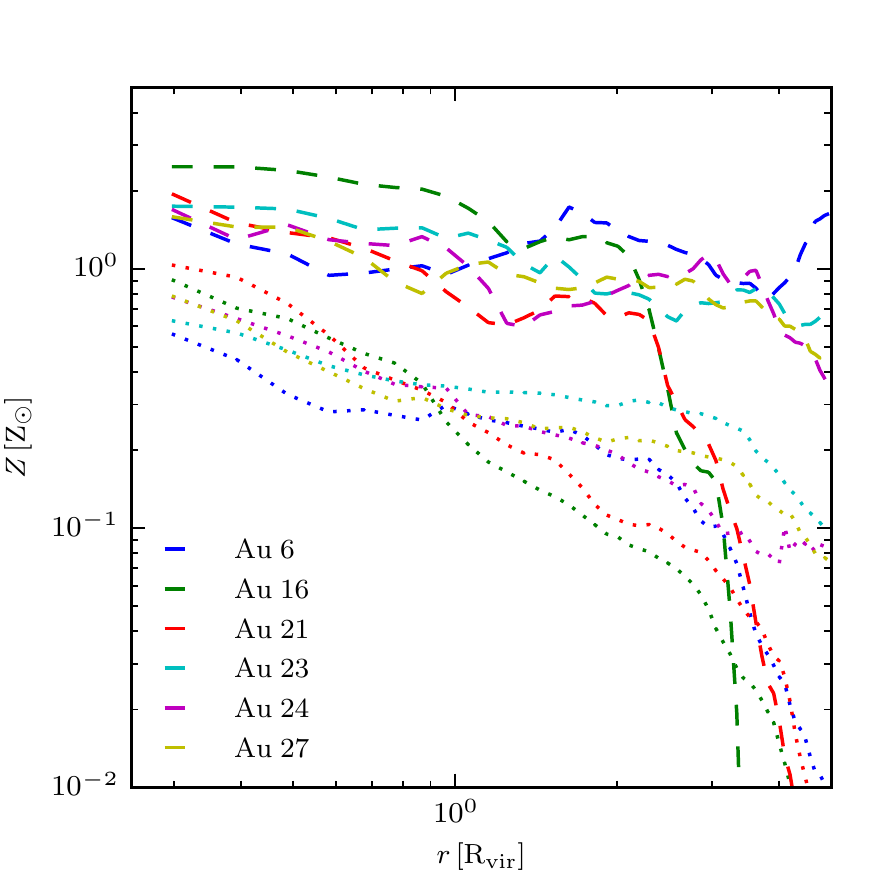}
  \caption{Radial profiles of the volume weighted root mean square magnetic field strength (top panel) and mass weighted metallicity (bottom panel) for the $6$ high resolution \textsc{Auriga} haloes at $z=0$ in units of the virial radius in a cone with an opening angle of $60^{\circ}$ around the $z$-axis (dashed lines), which is aligned with the angular momentum vector of the stellar disk to measure quantities preferentially in the outflow, and in a horizontal torus around the $x$-$y$-plane (dotted lines) with the same opening angle, to measure quantities preferentially in the inflow.}
  \label{fig:cones}
\end{figure}

\section{Variability between galaxies}
\label{sec:lvl3}

Besides studying Au-6-CGM in detail, it is important to look at the variation of the magnetic field in the CGM between different galaxies with different cosmic histories. We show radial profiles of magnetic field strength and plasma beta $\beta = p_\mathrm{thermal}/p_\mathrm{magnetic}$ at $z=0$ in Fig.~\ref{fig:profiles}. Similar to Au-6-CGM the CGM of all haloes is magnetised well beyond the virial radius. The radial profiles of the magnetic field strength are remarkably similar. There is a variation of less than a factor of two in the normalisation of the profile, and its slope is very similar (close to  $r^{-1.5}$) for all haloes out to twice the virial radius. Satellite galaxies show up as a clear peak in the azimuthally averaged profile, but only change the profile locally without any obvious effect on larger scales. Although their ISM magnetic field strength is much stronger than the CGM field of the host galaxy, the total magnetic energy stored in the gas of satellite galaxy is still small compared to the total magnetic energy in the CGM of the host galaxy. Thus, even if all gas was stripped from the satellite galaxy and mixed into the halo, its contribution to the total magnetic energy in the halo would still be subdominant compared to outflows from the central galaxy and in-situ dynamo amplification in the halo for our Milky Way-like galaxies. However, this may be different for more massive haloes, including clusters of galaxies.

The ratio between thermal pressure and magnetic pressure is approximately constant for radii between a quarter of and two times the virial radius at a value of about $10$ and quite similar for different galaxies. Consistent with Au-6-CGM, the magnetic field never reaches equipartition in the CGM. 

\begin{figure*}
  \centering
   \includegraphics[width=\linewidth]{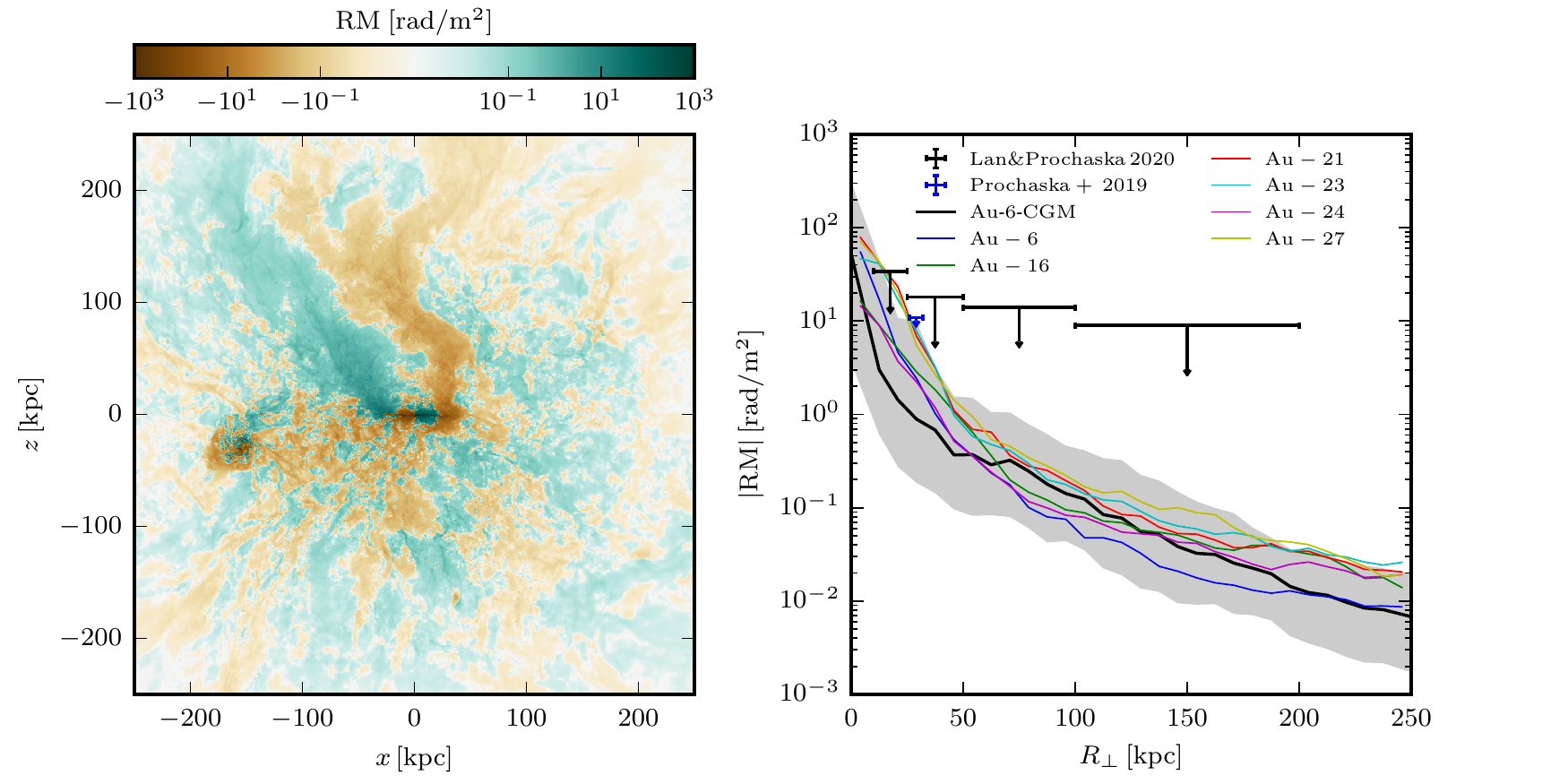}
  \caption{Edge-on map of the Faraday Rotation signal of Au-6-CGM with a minimum spatial resolution of $1\,\mathrm{kpc}$ (left panel). The median of the RM signal (right panel) are shown for Au-6-CGM as well as the high resolution \textsc{Auriga} haloes. The right panel also shows the $1\,\sigma$ percentiles for Au-6-CGM (dark shaded area) and recent upper limits by \citet{Prochaska2019} and \citet{Lan2020}. Substructure has \textit{not} been removed for this plot.}
  \label{fig:faraday}
\end{figure*}

To quantify the effect of outflows on the magnetic field in the CGM at low redshift we show the radial profiles of the average magnetic field strength and average metallicity in a cone around the $z$-axis that mostly contains the galactic wind. We compare them to profiles in a horizontal torus with the same opening angle around the plane of the disk that is mostly devoid of the galactic wind in Fig.~\ref{fig:cones}. The magnetic field strength is higher in the galactic wind compared to the part of the CGM that is not directly affected by the wind, but only by about a factor of $\approx 2$. In contrast, the metallicity in the outflow is about an order of magnitude higher in the wind compared to the wind-free part of the CGM. If magnetised outflows were the dominant path to magnetise the CGM at low redshift we would expect the contrast of the magnetic field strength in the outflow with respect to other parts of the CGM to be larger than the difference in the metallicity, as numerical dissipation will cause magnetic fields to decay over time while metals accumulate. Note however, that the physical diffusivity in the CGM is likely smaller than the numerical diffusivity of our code. The opposite result, in contrast, is another strong sign of an in-situ dynamo operating in the CGM that sets the strength of the magnetic field, consistent with our detailed analysis of Au-6-CGM.

\section{Synthetic Faraday rotation maps}
\label{sec:faraday}

As just very recently shown by \citet{Prochaska2019} it is possible to use fast radio bursts to measure Faraday rotation in the halo of galaxies. They measure a Faraday rotation measure (RM) value of $\approx 10\,\mathrm{rad}\,m^{-2}$ at a distance of $30\,\mathrm{kpc}$ to a galaxy that has a stellar mass very similar to the Milky Way.

In Fig.~\ref{fig:faraday} we show an edge-on Faraday rotation map of Au-6-CGM and profiles of the Faraday rotation signal and its $1\sigma$ percentiles for Au-6-CGM and the 6 high resolution \textsc{Auriga} galaxies. They are computed in the same way as in \citet{Pakmor2018}. As the map shows, the signal reverses sign on scales of several $10\,\mathrm{kpc}$ in most regions, though there are more coherent regions in coherent outflows, mirroring the structure of the magnetic field. The strength of the RM signal declines steeply with radius as both the thermal electron density and the magnetic field strength decline. There is significant scatter between different lines of sight, not only for different galaxies, but also for different lines of sight through the CGM of the same galaxy. The median at a given impact parameter varies by up to an order of magnitude between the different galaxies. Moreover the upper $1\sigma$ percentile is more than an order of magnitude bigger than the lower $1\sigma$ for the individual galaxies at a given impact parameter.

The upper limit of $\approx 10\,\mathrm{rad/m^2}$ at an impact parameter of $30\,\mathrm{kpc}$ measured by \citet{Prochaska2019} is completely consistent with our simulations. Their conclusion that the magnetic field strength in the CGM is significantly below equipartition, as discussed in Sec.~\ref{sec:lvl3}, is confirmed in our simulations as well. The structure of the magnetic field at the transition between disk and CGM seems to be consistent with the inferred structure of M51 \citep{Kierdorf2020}.

More recently, \citet{Lan2020} argue that they obtain upper limits on the RM in the CGM for distances up to $100\,\mathrm{kpc}$.  These limits are also consistent with our medians and most lines of sight. Note that the sample used by \citet{Lan2020} consists mostly of galaxies that are less massive than the sample we look at here.

Unfortunately, however, the typical strength of the RM signal at a radius of $100\,\mathrm{kpc}$ is already two orders of magnitude smaller than at $30\,\mathrm{kpc}$, making it generally very challenging to observe in the near future.

\section{Summary, discussion and outlook}
\label{sec:conclusion}

In this paper we analysed the high resolution simulations of the \textsc{Auriga} project and additional re-simulations with extra uniform resolution in the CGM (Au-6-CGM) to understand the evolution of the magnetic field in the CGM of Milky Way-like disk galaxies. We find that there are two important processes that shape the magnetic field in the CGM. At high redshift outflows of magnetised gas from the disk increase the magnetic field strength in the CGM, as can be seen from the comparison of the spatial distribution of metallicity and magnetic field strength shown in Fig.~\ref{fig:proj_au6_bfld} to Fig.~\ref{fig:bfldmetals}. The resulting magnetic field in the CGM is a chaotic small-scale field.

At low redshift, an in-situ turbulent dynamo in the halo further amplifies the small-scale field that originated from outflows from the disk and pushes magnetic energy to larger scales. This turbulent dynamo operates on timescales of Gigayears and saturates when the magnetic energy reaches about $10\%$ of the kinetic energy in the halo, i.e. well below equipartition, as seen in the evolution of the magnetic and kinetic power spectra in Fig.~\ref{fig:powerspectra} and the radial profiles of different energy densities in Fig.~\ref{fig:profilestime}.

We show that the results of our analysis of Au-6-CGM are consistent with all high resolution galaxies of the \textsc{Auriga} project. The variation between haloes is relatively small (see Fig. 9 - 11). Finally, we compare synthetic Faraday rotation maps of the CGM of our simulations with recent observations \citet{Prochaska2019} and find excellent agreement (see Fig.~\ref{fig:faraday}).

Our results show qualitative similarities to earlier simulations of magnetic fields in the halo of Milky Way-like galaxies \citep{Beck2012}. Similar to our simulations, they found that the halo is filled with a magnetic field. However, our results show significant differences. At $z=0$, the magnetic field at the virial radius is much stronger ($\approx 0.1\mu G$) in our simulations compared to a field strength of $\approx 10^{-3}\mu G$ at the virial radius in \citet{Beck2012}. Moreover, the radial profile of the magnetic field strength at $z=1$ or $z=0$ does not show any break in their simulations out to at least $1\,\mathrm{Mpc}$. We generally associate these differences to very different feedback models (e.g. \citealt{Beck2012} used much weaker feedback than needed to form realistic disk galaxies) and the more accurate numerical scheme we employ that allows us, together with advances in computing power, to simulate the CGM at much higher spatial resolution and with better accuracy.

Interestingly, also the magnetic field in the cosmological zoom-simulations of the \textsc{Fire} project have magnetic fields in the disk and the halo that are significantly (by about a factor of $10$) smaller than the magnetic fields we find or that are observed for the Milky Way \citep{Hopkins2019}. Their different magnetic field is likely a result of a different ISM and feedback model and a more diffusive numerical scheme.

A highly magnetised CGM as found in our simulations and consistent with very recent observations \citep{Prochaska2019}, has interesting consequences for future observations. Because the magnetic field strength is surprisingly large even at the virial radius ($B\approx 0.1\,\mu G$) it should in principle be hard but possible to detect those fields. Nevertheless, the densities of thermal and cosmic ray electrons in the CGM are still significantly lower than in the disk, so detecting a magnetic field in the CGM of Milky Way-like galaxies at galactocentric distances of several $10\,\mathrm{kpc}$ or beyond remains very challenging. Moreover, because the magnetic field strength varies significantly on the smallest scales in the CGM, at a given radius there will be significant scatter for observables that trace thin lines of sight through the CGM, such as Faraday rotation measurements of bright polarised background sources.

For the future, we need more high resolution CGM simulations of galaxies with a larger range of halo masses to better understand which of our results are specific to Milky Way-like systems. Moreover, the whole picture could still change once additional physical processes like cosmic rays or thermal conduction are included, that are neglected in CGM simulations so far \citep{Buck2020}.

\section*{Data availability}
The simulations underlying this article will be shared on reasonable request to the corresponding author.

\section*{Acknowledgements}
We thank the anonymous referee for interesting and detailed comments that significantly improved the quality of this paper. FvdV was supported by the Deutsche Forschungsgemeinschaft through project SP 709/5-1. FAG acknowledges financial support from CONICYT through the project FONDECYT Regular Nr. 1181264, and funding from the Max Planck Society through a Partner Group grant. TG acknowledges funding by European Research Council through the ERC advanced grant No. 787361-COBOM. FM acknowledges support through the Program "Rita Levi Montalcini" of the Italian MIUR. CP acknowledges support acknowledge support by the European Research Council under ERC-CoG grant CRAGSMAN-646955. This research was supported in part by the National Science Foundation under Grant No. NSF PHY-1748958.
    
\bibliographystyle{mnras}

\label{lastpage}

\end{document}